\documentclass[lettersize,journal]{IEEEtran}
\usepackage{amsmath,amsfonts}
\usepackage{algorithmic}
\usepackage{algorithm}
\usepackage{array}
\usepackage[caption=false,font=normalsize,labelfont=sf,textfont=sf]{subfig}
\usepackage{textcomp}
\usepackage{stfloats}
\usepackage{url}
\usepackage{verbatim}
\usepackage{graphicx}
\usepackage{cite}
\usepackage{url}
\usepackage{soul}
\usepackage{makecell}

\usepackage{multirow}
\usepackage{subfiles}
\usepackage{booktabs}
\usepackage{flafter}
\newcommand\figwidth{0.417\textwidth}

\begin{document}
\bstctlcite{IEEEexample:BSTcontrol}

\title{Map-Based Path Loss Prediction in Multiple Cities Using Convolutional Neural Networks}

\author{Ryan G. Dempsey, Jonathan Ethier, and Halim Yanikomeroglu
\thanks{R. G. Dempsey and J. Ethier are with Communications Research Centre Canada (CRC), Ottawa, ON, Canada. R. G. Dempsey and H. Yanikomeroglu are with Carleton University, Ottawa, ON, Canada. Corresponding author: Ryan G. Dempsey (email: ryan.dempsey@ised-isde.gc.ca).}
}


\maketitle

\begin{abstract}
Radio deployments and spectrum planning benefit from path loss predictions. Obstructions along a communications link are often considered implicitly or through derived metrics such as representative clutter height or total obstruction depth. In this paper, we propose a path-specific path loss prediction method that uses convolutional neural networks to automatically perform feature extraction from 2-D obstruction height maps. Our methods result in low prediction error in a variety of environments without requiring derived metrics.
\end{abstract}


\section{Introduction}
\IEEEPARstart{P}{ath} loss modeling is a tool used by the wireless industry and regulatory bodies to optimize wireless coverage and avoid unwanted interference. Path loss models predict point-to-point (P2P) wireless signal attenuation along a communications link from the transmitter (Tx) to the receiver (Rx). Path loss is primarily affected by frequency, link distance, and obstructions in the line of sight (LOS). In LOS conditions, free space path loss (FSPL), which uses frequency and distance, is sufficient for accurate P2P path loss predictions. In non-LOS conditions, however, FSPL typically underestimates path loss.

LOS obstructions are identified from surface-level variations, captured by the geographic information system (GIS) quantity digital surface model (DSM). DSM captures heights relative to sea level: ground-level variations (terrain) as well as trees, buildings, and other structures (clutter). Where there is no clutter, DSM models the terrain.

Traditional path loss models, such as Longley-Rice \cite{longleyrice}, explicitly account for terrain information but do not explicitly consider the presence of clutter. Consequently, these predictions can underestimate path loss in links dominated by clutter. Conversely, modern path loss models that consider clutter information have significantly lower prediction errors in non-LOS links. One such model, P.1812-6 \cite{p1812} (recommended by the International Telecommunications Union Radiocommunication Sector (ITU-R)), considers clutter along the link in the form of representative clutter heights. The heights are calculated based on local clutter types, improving prediction accuracy but remaining limited by approximations in clutter height estimation and focusing on diffraction losses \cite{feature-pl}.

Previous work in \cite{feature-pl} shows improvement over P.1812-6 in non-LOS conditions, by capturing GIS information in one novel feature: total obstruction depth. This scalar feature is computed using surface information along the direct path. Alongside frequency and distance, obstruction depth yields a root mean square error (RMSE) in the range of 6-8~dB as compared to 8-13~dB for P.1812-6 \cite{feature-pl}. While using total obstruction depth as a feature can produce low errors, it does not explicitly consider the surface height profile, which can influence path loss in more complicated ways due to diffraction and Fresnel zone effects \cite{fresnel}. Designing and computing engineered features can also be time-consuming or difficult.

The goal of this paper is to explicitly utilize the surface height profile to train a path-specific propagation model akin to P.1812 (unlike a model trained only on derived features \cite{feature-pl}). Rather than aiming to reduce RMSE compared to \cite{feature-pl}, our objective is to study whether deep neural networks can automatically extract spatial features from GIS data to streamline the process over feature engineering. Leveraging high-resolution clutter maps and deep learning to directly predict path loss, this paper will determine whether automatic feature extraction and regression integrated in a single neural network can achieve RMSE lower than P.1812 and similar to \cite{feature-pl}.

Previous work includes \cite{physics-cnn-indoor, urban-prediction, cnn-satellite}, which demonstrate generalized path loss prediction in spatial holdouts, using simulated data in \cite{physics-cnn-indoor} or measurement data in \cite{urban-prediction, cnn-satellite}. Conducting geographic cross-validation within a single region, \cite{manhattan-transactions} provides a robust assessment of generalization using measurement data. Similarly, \cite{gis-transactions} uses map-based inputs to train path loss models limited to two frequencies in a single region. Our methods use geographic cross-validation across multiple cities--a prerequisite to generalized models that can accurately predict in new regions \cite{gis-split}. Our work differs from \cite{physics-cnn-indoor, urban-prediction} by training and testing exclusively on measurements, from \cite{feature-pl, physics-cnn-indoor, urban-prediction} by explicitly using the surface height profile as an input, and from \cite{physics-cnn-indoor, urban-prediction, cnn-satellite, manhattan-transactions, gis-transactions}, by conducting geographic cross-validation across multiple cities. To our knowledge, the novel combination of explicitly using the surface height profile, exclusively training and testing on measurements, and evaluating with geographic cross-validation across multiple cities has not been explored.

This paper describes path loss models trained on 2-D obstruction height maps alongside measurement data. Section~\ref{data-prep} describes the preparation of training and evaluation data. Section~\ref{feature-arch} describes the 2-D features and the model architecture. Section~\ref{train} describes training and evaluation, and Section~\ref{conc} summarizes our findings.

\section{Data Preparation}
\label{data-prep}
\subsection{Dataset Description}
In this paper, we use two datasets: radio measurement data, and GIS elevation/clutter data. For radio measurement data, we use drive test measurement data from the United Kingdom Office of Communications (UK Ofcom) \cite{uk-ofcom}, which can be accessed by the general public through Ofcom's open data repository \cite{open-uk}. This dataset was collected between 2015 and 2018, and provides 8.2 million measurements across six frequency bands (449, 915, 1802, 2695, 3602, and 5850 MHz) in seven cities in the UK. Each city uses a single Tx with varying Rx locations. Link distances vary from 1 m to 78 km, with a median of 8 km.

For clutter data, we use the UK open digital elevation model (DEM) \cite{uk-dsm-dtm}. This dataset provides DSM at a 1~m resolution for various areas across the UK in 2021. DEM data is available for six out of seven drive test cities: Boston, London, Merthyr Tydfil, Nottingham, Southampton, and Stevenage. We use data from these six cities to train and test our path loss models.


\subsection{Data Cleaning}
\subsubsection{Ofcom drive test data}
Firstly, we use measurements with received signal level (\(RSL\)) above the system noise floor provided in the dataset, by a 6 dB margin, as shown in \eqref{noisefloor}. Similarly, we use only measurements with a link distance \(d\) above 50~meters as in \eqref{distancefilter}:
\begin{subequations}
\label{filtering}
\begin{eqnarray}
    \label{noisefloor}
    {RSL}\ [\mathrm{dBm}] > \mathrm{noise\ floor}\ [\mathrm{dBm}] + 6\ \mathrm{dB}, \\
    \label{distancefilter}
    d\ [\mathrm{m}] > 50\ \mathrm{m}.
\end{eqnarray}
\end{subequations}
This allows training and evaluation to be conducted only on samples that have a reasonable expectation (and benefit) to be modeled. After both filtering steps, 4.2 million measurements remain for joining with DSM data. Link distances in this subset vary from 50~m to 57~km, with a median of 7~km.

\subsubsection{Digital surface model data}
We first set negative DSM values to 0, as they are likely GIS measurement errors. To extract path profiles from the data, we generate a rectangular grid of width \(W\ [\mathrm{m}]\) at a 1~m resolution in the projected coordinate system in the UK. Orienting the Tx and Rx locations in the center of the transverse axis, we create an array of \((\left \lfloor{d} \right \rfloor \times W)\) sets of evenly-spaced coordinates (where \(\left \lfloor\ \right \rfloor\) denotes the floor function). We choose \(W\) to be an odd integer, allowing the Tx and Rx locations to reside on the closest single pixel each, and enabling the creation of a ``direct path'' feature consisting of a single line in the center of the path. This feature is described in Section \ref{feature-arch}. The final path profiles are extracted using 2-D nearest neighbor interpolation \cite{splines} of the DSM data at the generated grid points. This results in a 2-D DSM array of shape \((\left \lfloor{d} \right \rfloor \times W)\), where the orientation between Tx and Rx in each array is held constant. We set \(W=61\ \mathrm{m}\) to cover the majority of the first Fresnel zones for these frequencies while having a practical computational load. Finally, we apply earth curvature correction \cite[eq. (7-16)]{spherical-earth} to the DSM heights relative to the Tx, using a radius of 6365~km.

\section{Feature and Architecture Selection}
\label{feature-arch}
We formulate this work as an image problem, where each input feature forms a channel \cite{deeplearning-ref} of a 2-D image.

\subsection{Feature Selection}
Before computing the feature channels, we re-sample all images to have a direct path length of 256 pixels, using bilinear interpolation \cite{bilinear} . While the majority of path profiles are down-sampled, some of the shorter links result in up-sampling to 256. Initial experimentation with variably-sized array lengths proportional to the corresponding link distances yielded poor results with significantly more computational requirements. Re-sampling to a uniform size allows for the design of a single architecture that generalizes well for all link distances. We choose a resolution of 256 pixels for its consistently low validation RMSE during architecture selection, with higher resolutions suffering from overfitting.

In the re-sampled arrays, we include the essential features, frequency, and distance, as their own channels. Frequency~[MHz] is encoded as a channel with a constant value, while the distance channel is generated by computing the 2-D Euclidean distance [m] from the Tx to each pixel. The distance channel serves the dual purpose of encoding relative spatial locations \cite{coordconv} while providing the non-re-sampled distances.

We use two additional channels to capture path profile information: (1) DSM and (2) direct path. DSM provides the path profile heights and direct path provides the antenna locations and heights, and the 3D path from Tx to Rx. Antenna heights relative to sea level are obtained by adding the heights provided by Ofcom to the terrain. This channel contains the direct path heights in the center of the transverse dimension and 0 elsewhere.

\begin{figure}[!t]
\centering
\captionsetup[subfloat]{font={default}, labelfont={default}, textfont={default}}
\subfloat[]{\centering \includegraphics[width=\figwidth]{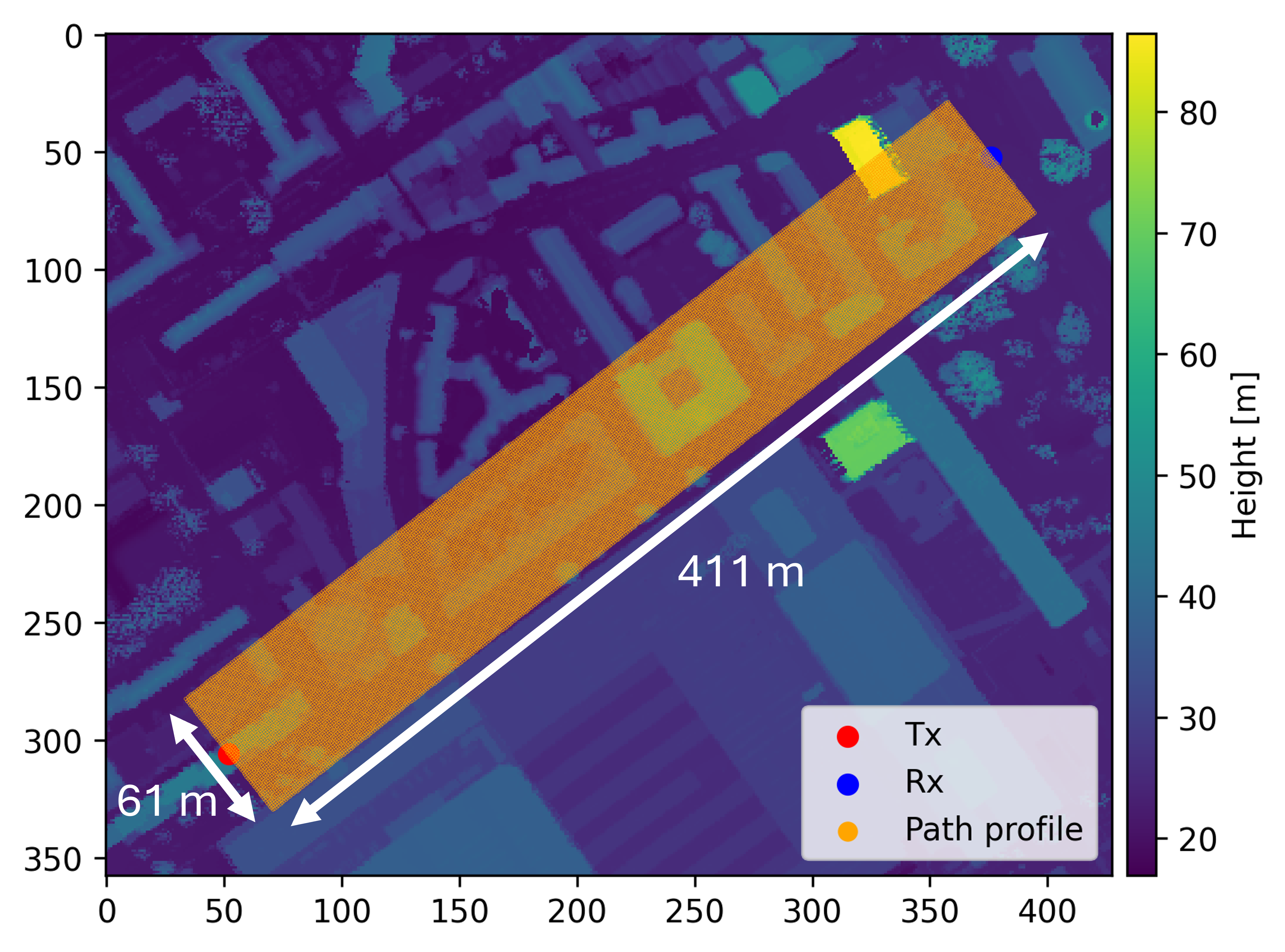}
\label{rastersample}}
\hfil
\subfloat[]{\centering \includegraphics[width=\figwidth]{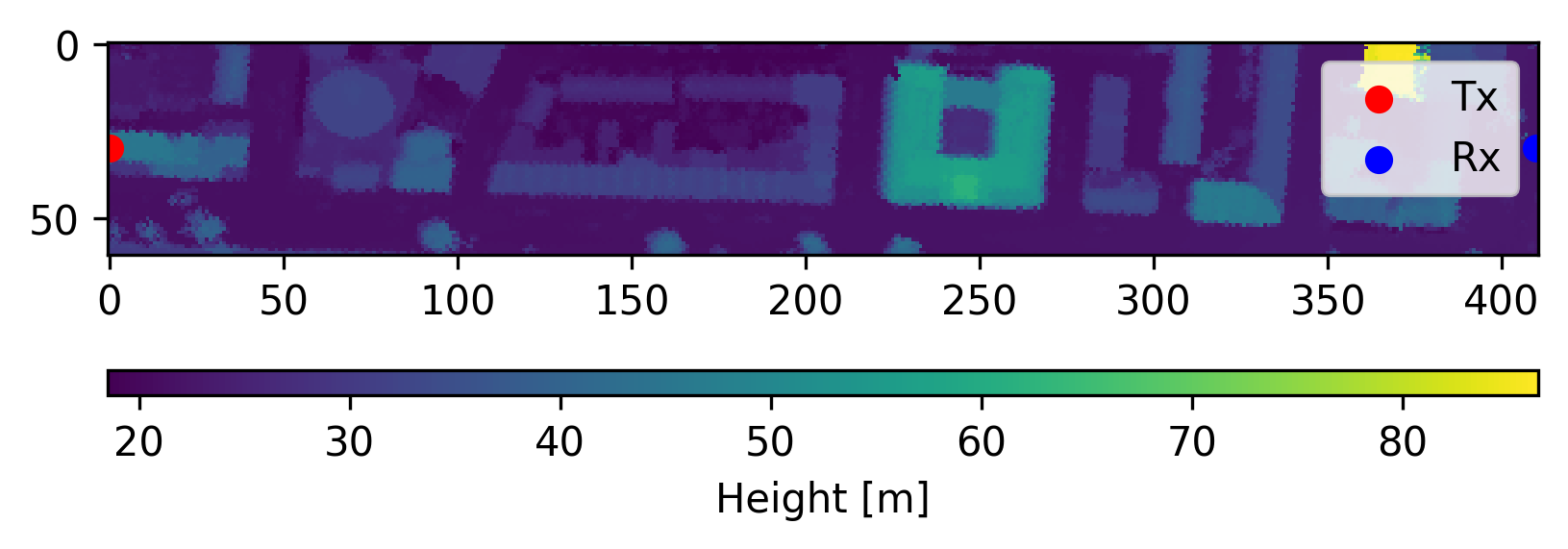}
\label{pathprofilefull2d}}
\caption{Path profile extraction process for a link in London. Clutter presents as higher intensity, while roads present as lower intensity. (a) Subset of raster used for path profile extraction. (b) Extracted path profile (not normalized).}
\label{path-profile-extraction}
\end{figure}

Fig.~\ref{path-profile-extraction} shows the path profile extraction process and result for a link in London with a distance of 411~m. The final dataset contains samples of shape:\
\begin{math}\\(4~\mathrm{channels} \times 256~\mathrm{length}\times W~\mathrm{width})\end{math}, where \(W=61\).

All features and their normalizations are summarized in Table~\ref{feature-table}. Locality refers to where the normalization is retrieved from. All heights are normalized on a per-sample basis. While the model is able to exploit the frequency dependence of the features in the transverse dimension, we note that normalizing the heights for each sample prevents the model from fully exploiting frequency dependence along the height dimension, by not having access to the scale of the heights. Attempts to include height scaling within the input yielded significant overfitting. Frequency dependence is fully exploitable by the model along the transverse dimension, since this dimension is not re-sampled, and distances and frequencies are both present.

\begin{table}[!t]
\centering
\caption{Input Features and Normalizations\label{feature-table}}
    \begin{tabular}{p{2.2cm}p{4cm}p{1.24cm}}
    \toprule
        Channel & Scaling & Locality \\
        \midrule
        Frequency & Divided by 8000 MHz maximum & Globally \\
        \midrule
        Distance from Tx to each pixel & Divided by maximum link distance in the dataset & Globally \\
        \midrule
        Surface & \multirow{2}{*}{\makecell[l]{Normalized to [0, 1] using\\heights of surface and direct path}} & \multirow{2}{*}{Per sample} \\
        \cmidrule{1-1}
        Height of direct path from Tx to Rx & & \\
    \bottomrule
    \end{tabular}
\end{table}

\begin{figure}[!t]
\centering
\includegraphics[width=\figwidth]{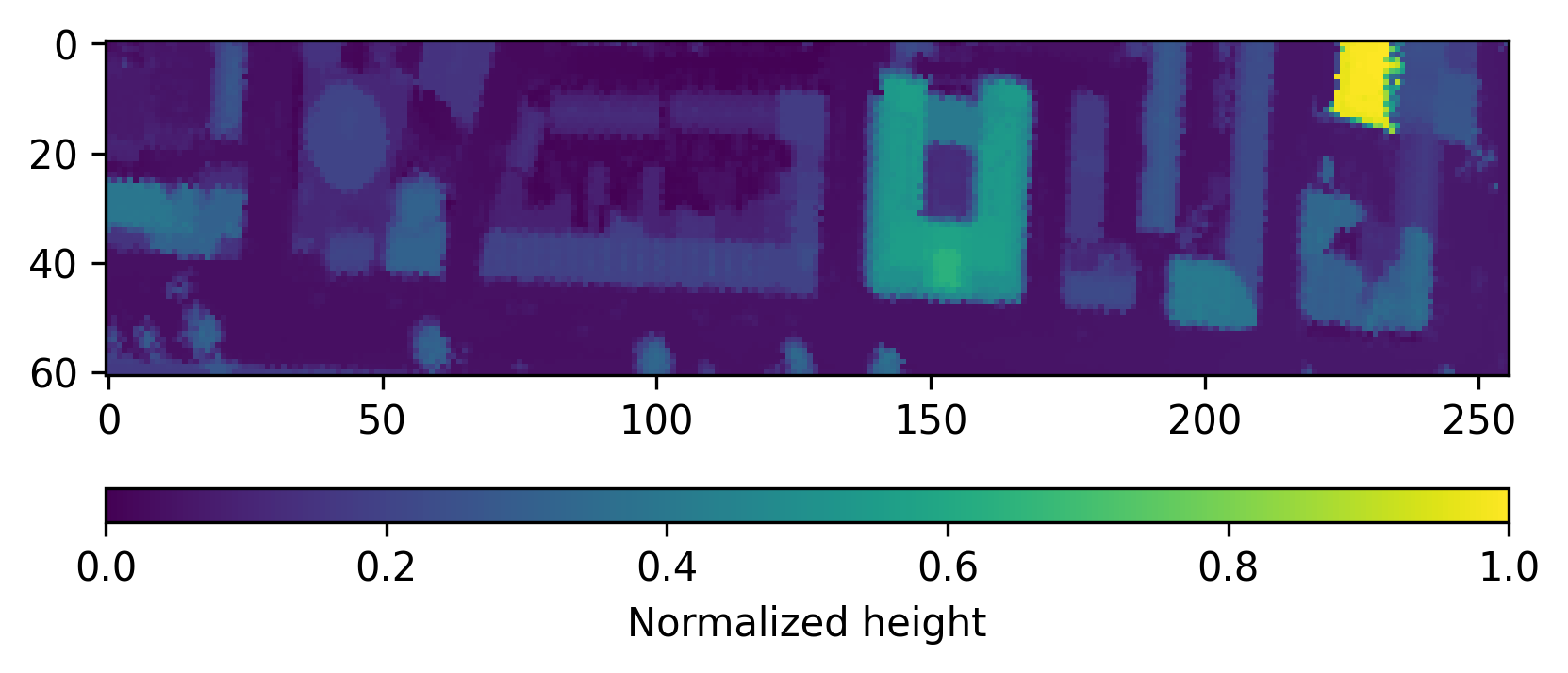}
\caption{Re-sampled and normalized surface channel input to CNN.}
\label{path-profile-normed-2d}
\end{figure}

\begin{figure}[!t]
\centering
\includegraphics[width=\figwidth]{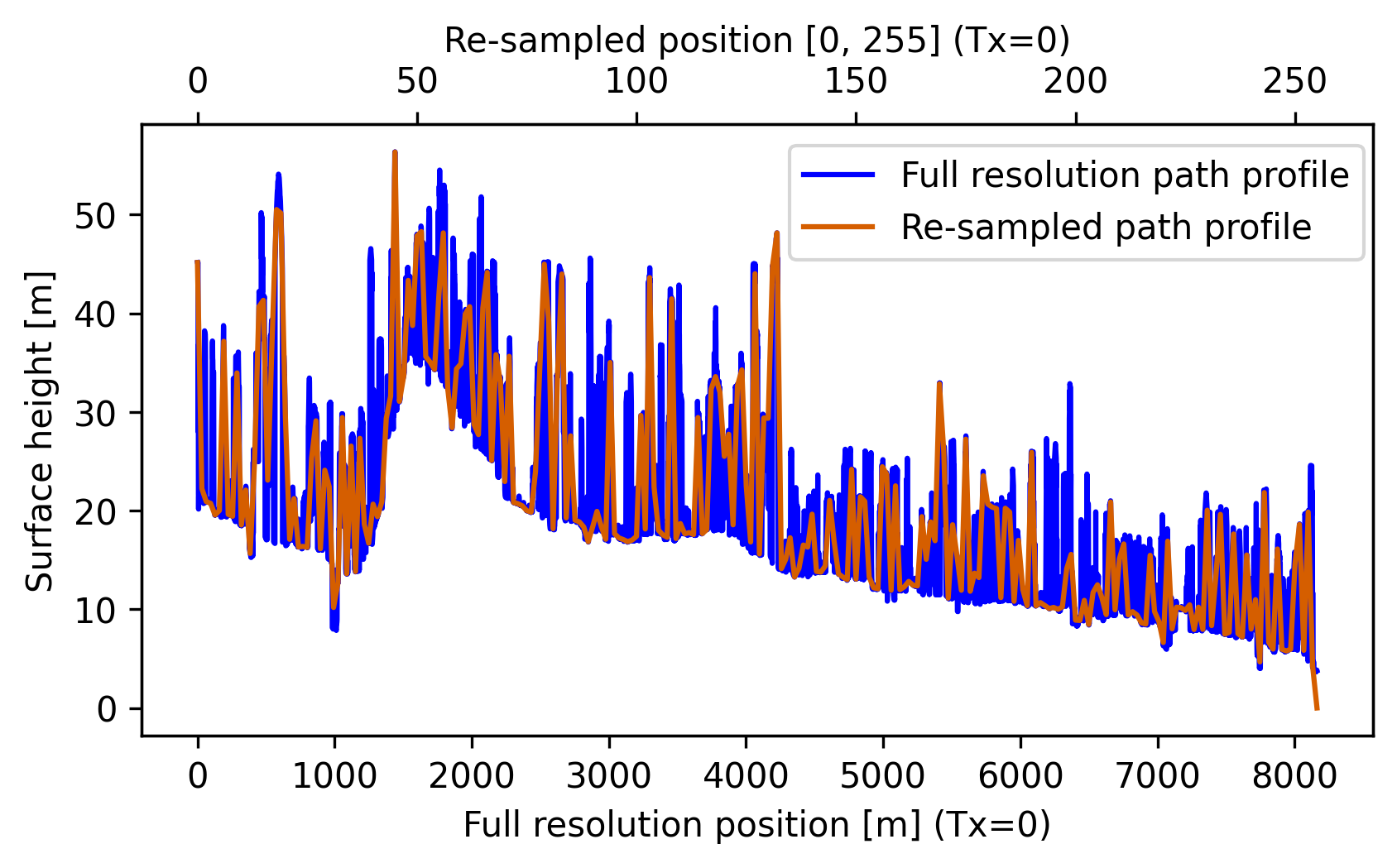}
\caption{Full size and re-sampled center of path profile for a link in London.}
\label{path-profile-comparison}
\end{figure}

Fig.~\ref{path-profile-normed-2d} shows a re-sampled and normalized input to the path loss model for the same link shown in Fig.~\ref{path-profile-extraction}. We also compare the effect of re-sampling on a typical link, showing both re-sampled and full DSM values for the center (direct path) of an 8~km in Fig.~\ref{path-profile-comparison}. Features such as thin building peaks are filtered out, lowering precision. However, the aforementioned validation improvements suggest that re-sampling provides a regularization effect.


\subsection{Model Architecture}
Given the spatial structure of the data, we employ convolutional neural networks (CNNs) \cite{deeplearning-ref}, treating the data as images. Each convolution layer has a kernel size of 3$\times$3, stride size of 2$\times$2, and zero-padding size of 1$\times$1. Each max pool layer has a kernel size of 3$\times$3, stride size of 1$\times$1, and ``same'' zero-padding \cite{deeplearning-ref}. During training, we use dropout at a rate of 25\% on the 4096 flattened features to minimize overfitting. The extracted features are input to a fully connected network (FCN) \cite{deeplearning-ref} with 256 hidden units and rectified linear unit (ReLU) activation. The final output is path loss. Given the low RMSE in \cite{feature-pl} with only 3 features, 4096 features are likely to be sufficient, regardless of the feature extraction process.

\begin{table}[!t]
\centering
\caption{Summary of CNN Architecture}
\label{full-conv-table-new}
    \begin{tabular}{p{3.89cm}p{4cm}}
    \toprule
    Layer & Output image shape\newline(channels \(\times\) length \(\times\) width)\\
    \midrule
    
    Input & 4 \(\times\) 256 \(\times\) 61\\
    \midrule
    
    Convolution + ReLU $\xrightarrow{}$ Max pool & 32 \(\times\) 256 \(\times\) 61 $\xrightarrow{}$ 32 \(\times\) 128 \(\times\) 31\\

    Convolution + ReLU $\xrightarrow{}$ Max pool & 64 \(\times\) 128 \(\times\) 31 $\xrightarrow{}$ 64 \(\times\) 64 \(\times\) 16\\

    Convolution + ReLU $\xrightarrow{}$ Max pool & 128 \(\times\) 64 \(\times\) 16 $\xrightarrow{}$ 128 \(\times\) 32 \(\times\) 8\\

    Convolution + ReLU $\xrightarrow{}$ Max pool & 256 \(\times\) 32 \(\times\) 8 $\xrightarrow{}$ 256 \(\times\) 16 \(\times\) 4\\

    Convolution + ReLU $\xrightarrow{}$ Max pool & 512 \(\times\) 16 \(\times\) 4 $\xrightarrow{}$ 512 \(\times\) 8 \(\times\) 2\\

    Convolution + ReLU $\xrightarrow{}$ Max pool & 1024 \(\times\) 8 \(\times\) 2 $\xrightarrow{}$ 1024 \(\times\) 4 \(\times\) 1\\
    \midrule
    Flatten + Dropout @ 0.25 & 4096\\
    Fully connected + ReLU & 256\\
    Fully connected & 1\\
    \bottomrule
    \end{tabular}
\end{table}

    
    





The network structure is summarized in Table~\ref{full-conv-table-new}. The model has 7\,337\,569 total parameters. Our CNNs are developed, trained, and evaluated in PyTorch \cite{torch}. The following section describes the training regimen and evaluation results.

\section{Training and Results}
\label{train}
We randomly sample 10\,000 unique links from each frequency/city combination, resulting in 60\,000 samples per city, for a total of 360\,000 samples. This ensures a uniform training distribution across different cities and frequency bands.

We use a six-city cross-validation, where each city is held out as the test set, with the remaining five cities split between train and validate. We use adaptive moment estimation (Adam) \cite{adam}, mean squared error loss, a learning rate of 0.0001, and a batch size of 256 for 200 epochs. The model with the lowest validation RMSE is used to evaluate test RMSE for each run.

\subsection{Validation Split}
Deep learning models are more likely to overfit to GIS arrays (feature-rich) than scalar features (feature-sparse) for the same amount of data, given the difference in dimensionality \cite{overfitting}. To mitigate this, for each city holdout, we geographically split each of the remaining five cities using 80\% for training and 20\% for validation. In each of these five cities, an initial angle \(\theta\) is chosen. Rotating counter-clockwise about the Tx, the first 20\% of Rx locations encountered are used as validation samples, with the remaining 80\% used for training. A geographic split for each of the five non-holdout cities provides diverse yet geographically distinct training and validation sets. This split is necessary with this feature set, as initial studies using random splits caused significant overfitting due to high correlation with training sets. 


\subsection{Evaluation of Result Consistency}
Each city holdout is conducted 10 times with different random seeds to evaluate model robustness. For each training run, we sample a 20\% subset (2000 samples) from each city/frequency combination from the original 10\,000 subset, for a total of 72\,000 samples (48\,000 training, 12\,000 validation, and 12\,000 test). Each run has a random 20\% subset, initial weights, training data shuffling, and initial validation angle \(\theta\) for each city.

Table~\ref{results-combined} summarizes the cross-validation results with the mean and standard deviation (SD) of test RMSE for each holdout. We also compare the means with ITU-R P.1812-6 \cite{p1812} (the industry standard for 1-6~GHz) and a three-feature FCN using frequency, distance, and total obstruction depth \cite{feature-pl}.

\begin{table}[!t]
\centering
\caption{Cross-Validation Test RMSE [\textnormal{dB}] Across 10 Runs}
\label{results-combined}
\begin{tabular}{lcccc}
\toprule
        & \multicolumn{2}{c}{CNN} & \multicolumn{2}{c}{Benchmarks \cite{feature-pl}} \\
       Holdout &            Mean & SD & P.1812 & FCN\\
\midrule
        Boston &       \textbf{7.18} & 0.27 & 11.4 & 7.56 \\
        London &       7.75 & 0.47 & 8.8 & \textbf{7.44} \\
Merthyr Tydfil &       8.07 & 0.39 & 13.4 & \textbf{7.22} \\
    Nottingham &       \textbf{6.99} & 0.23 & 12.6 & 7.07 \\
   Southampton &       \textbf{6.35} & 0.42 & 9.5 & 6.64 \\
     Stevenage &       \textbf{7.73} & 0.21 & 12.3 & 8.81 \\
\midrule
          Mean &       \textbf{7.35} & 0.33 & 11.3 & 7.46 \\
\bottomrule
\end{tabular}
\end{table}


The model RMSE increases monotonically with frequency, showing 6.31 and 8.15~dB on 449 and 5850~MHz respectively. RMSE also increases with distance, with the largest errors occurring on the extremely long links. Across all runs and holdouts, the CNN achieves 7.37 dB RMSE on the shortest 99.9\% of unique holdout links ($<$45~km), and 12.04 dB RMSE on the remaining 0.1\% ($<$800 links). This implies that re-sampling can have a significant effect on performance, especially for longer links. Future work should ensure more representation of long links in the training set, or only operate the model on shorter links to ensure confidence in prediction.

Compared to the FCN, our CNN shows lower mean RMSE on four of six holdouts, with higher mean RMSE on London and Merthyr Tydfil. Though these trends may differ when attempting to generalize to other countries and measurements, using multi-city cross-validation suggests that the result would likely generalize to other unseen regions. Neither model provides strictly lower RMSE, with similar aggregate results. The choice of model depends on the user's priorities: The CNN provides the novelty of streamlining modeling by eliminating feature engineering but requires a larger initial investment in training time--80 minutes per run vs. 2 minutes for the FCN. The inference speed of the CNN and the FCN are 4000 and 10\,000 samples/s, respectively, with the same hardware (Amazon Web Services g5.4xlarge), which are both practical speeds. Finally, the fact that distinct modeling approaches applied to the same DSM achieve similarly low RMSE validates the use of DSM in machine learning-based path loss models, whether features are extracted automatically (CNN) or manually (FCN).

The CNN showing similar results to the FCN with richer inputs suggests room for further model refinement to reduce overfitting (as is often the case in machine learning research). Still, our novel CNN-based approach provides significant improvement over the current industry standard, achieving lower RMSE than P.1812 in all six holdouts. This implies that higher-dimensional information outperforms representative clutter heights in predictive models.

Merthyr Tydfil, London, and Stevenage are the three highest holdout RMSEs. The London holdout had consistently high RMSE while testing several hyperparameter configurations. To mitigate this, we tuned our hyperparameters to minimize London RMSE (sequential overfitting \cite{sequential-overfit}). Choosing a hyperparameter configuration that brought London under 8~dB RMSE resulted in a trade-off where Stevenage RMSE was increased to a similar value to London. Despite this, London is still in the upper half of holdout RMSEs, with the other half showing strong generalization. Since London contains the highest clutter density among the six cities, and Merthyr Tydfil contains the highest terrain density, it is more challenging to generalize high-dimensional GIS features to these two holdouts. Stevenage is also difficult to generalize to because it is a relatively new suburb, with newer building materials and different electromagnetic behavior. Additionally, there is likely more rapid expansion occurring compared to the other holdouts, potentially making the three-year discrepancy between measurement and DSM data more significant. Test performance in all three holdouts may be improved with additional channels, such as a terrain channel to differentiate between terrain and clutter, or a building materials channel. Aligning the datasets in time is also likely to decrease RMSE.

Future studies will investigate additional feature channels to decrease test RMSE, such as height scaling and Fresnel information within the path profiles, terrain to distinguish between clutter and ground, and type of clutter. Future work may also study further validation splitting techniques to minimize overfitting. Finally, including DSM height information in the area behind the Tx and Rx antennas will likely reduce RMSE as a result of its impact on path loss.

\section{Conclusion}
\label{conc}
Path loss modeling is an asset for effective radio deployments and spectrum planning. This paper described the use of CNNs in conjunction with surface path profiles to train a path-specific path loss model with low test RMSE in a variety of environments. We showed that CNNs trained to provide automatic feature extraction from 2-D path profiles achieve similar RMSE to a modern FCN path loss model with scalar features. These CNNs consistently achieved lower RMSE than the industry standard in a variety of communications environments, without the need for derived clutter metrics or scalar GIS features.

\section*{Acknowledgment}
The authors of this paper would like to thank Mathieu Châteauvert from Communications Research Centre Canada (CRC) for his support with propagation and GIS, as well as Alexis Bose (CRC) and Paul Guinand (CRC) for their advice on machine learning strategies.

\bibliographystyle{IEEEtran}
\bibliography{references}
\vfill
\end{document}